\documentclass[showpacs,preprintnumbers,amsmath,amssymb,nofootinbib]{revtex4}

\usepackage{graphicx}
\usepackage{amssymb}

\def\simge{\mathrel{%
       \rlap{\raise 0.511ex \hbox{$>$}}{\lower 0.511ex \hbox{$\sim$}}}}
\def\simle{\mathrel{
       \rlap{\raise 0.511ex \hbox{$<$}}{\lower 0.511ex \hbox{$\sim$}}}}

\begin{document}


\title{Equation of state in 2+1 flavor QCD with improved Wilson quarks by the fixed scale approach} 


\author{T.~Umeda}\affiliation{Graduate School of Education, Hiroshima University, Hiroshima 739-8524, Japan}
\author{S.~Aoki}\affiliation{Graduate School of Pure and Applied Sciences,
University of Tsukuba, Tsukuba, Ibaraki 305-8571, Japan}
\affiliation{Center for Computational Physics, University of Tsukuba, Tsukuba, Ibaraki 305-8577, Japan}
\author{S.~Ejiri}\affiliation{Graduate School of Science and Technology, Niigata University, Niigata 950-2181, Japan}
\author{T.~Hatsuda}\affiliation{Department of Physics, The University of Tokyo,
Tokyo 113-0033, Japan}
\affiliation{IPMU, The University of Tokyo, Kashiwa 277-8583, Japan}
\affiliation{Theoretical Research Division, Nishina Center, RIKEN, Wako 351-0198, Japan}
\author{K.~Kanaya}\affiliation{Graduate School of Pure and Applied Sciences,
University of Tsukuba, Tsukuba, Ibaraki 305-8571, Japan}
\author{Y.~Maezawa}\affiliation{Physics Department, Brookhaven National Laboratory, Upton, New York 11973, USA
} 
\author{H.~Ohno\footnote{Current address: Fakult\"at f\"ur Physik, Universit\"at Bielefeld, D-33615 Bielefeld, Germany}}\affiliation{Graduate School of Pure and Applied Sciences, University of Tsukuba, Tsukuba, Ibaraki 305-8571, Japan}

\collaboration{WHOT-QCD Collaboration}

\date{\today}

\pacs{12.38.Gc,12.38.Mh}

\begin{abstract}
We study the equation of state in 2+1 flavor QCD 
with nonperturbatively improved Wilson quarks coupled with the RG-improved Iwasaki glue.  
We apply the $T$-integration method to nonperturbatively calculate the 
equation of state by the fixed-scale approach.
With the fixed-scale approach, we can purely vary the temperature on a line of constant physics without changing the system size and renormalization constants.
Unlike the conventional fixed-$N_t$ approach, 
it is easy to keep scaling violations small at low temperature in the fixed-scale approach.
We study 2+1 flavor QCD at light quark mass corresponding to $m_\pi/m_\rho \simeq 0.63$, while the strange quark mass is chosen around the physical point. 
Although the light quark masses are still heavier than the physical values, 
our equation of state is roughly consistent with recent results with highly improved staggered 
quarks at large $N_t$.
\end{abstract}

\maketitle

\section{Introduction}
\label{sect1}

The QCD equation of state (EOS) at high temperature plays a key role in understanding 
the nature of quark 
gluon plasma (QGP), e.g.\ as inputs of the hydrodynamical description 
of QGP space-time evolution in heavy-ion collision experiments \cite{Hirano:2008hy}.
Lattice QCD simulations provide us with the only systematic way to calculate the EOS nonperturbatively without resorting to phenomenological assumptions.

For a quantitatively reliable evaluation of EOS in QCD, it is indispensable to incorporate dynamical up, down, and strange quarks.
However, dynamical quarks require a large computational effort on the lattice.
Most calculations of EOS have been made in the fixed-$N_t$ approach, in which the temperature $T=(N_t a)^{-1}$ is varied on a lattice with fixed temporal size $N_t$ by varying the lattice spacing $a$ through a variation of coupling parameters on a line of constant physics (LCP). 
Here, we note that a sizable fraction of the total computational cost is required to systematically carry out zero-temperature simulations to determine the location of the LCP, to get basic information such as the scale  and beta functions on the LCP, and to renormalize finite-temperature observables such as the EOS at each simulation point.
In QCD with dynamical quarks, such systematic simulations are quite demanding.

We adopt the fixed-scale approach, in which we vary $T$ by varying $N_t$ at a fixed $a$ \cite{Levkova:2006gn,Umeda:2008bd}. 
In this approach, because all the simulations are done with the same values of the coupling parameters, they are automatically on the same LCP. 
Furthermore, we need zero-temperature simulation at only one point to renormalize the observables at all $T$'s.
Thus, the cost for the zero-temperature simulations can be largely reduced.
To take or to confirm the continuum limit, we may repeat the calculations at several values of $a$.
As the zero-temperature configurations, we may even borrow high statistic configurations on fine lattices, which were generated for spectrum studies at $T=0$ and are open to the public on the International Lattice Data Grid (ILDG) \cite{Maynard:2010wi}. 

The fixed-scale approach is complemental to the conventional fixed-$N_t$ approach in several respects:
In the very high $T$ region where $T \simge {\cal O}(a^{-1})$, the fixed-scale approach suffers from lattice artifacts due to the coarseness of the lattice in comparison with the typical extent $T^{-1}$ of thermal fluctuations, while in the fixed-$N_t$ approach one can keep $T^{-1}/a = N_t$ finite even in the high temperature limit.
In the fixed-scale approach, the spatial volume of the system is kept fixed 
at all $T$'s with the same spatial lattice size $N_s$, while in the 
fixed-$N_t$ approach the $N_s$ has to be increased to quite large values at 
high $T$'s to keep the spatial volume.
Large spatial volume is important at light quark masses to suppress volume effects in the hadron spectrum and thus in the determination of the scale and LCP. 
At small $T$'s, typically at $T \simle T_{\rm pc}$, where $T_{\rm pc}$ is the pseudocritical temperature, the fixed-scale approach keeps a small $a$,  
while the fixed-$N_t$ approach suffers from lattice artifacts due to large $a$.
It should be kept in mind here that the fixed-scale approach requires high statistics in the low $T$ region, where we have a severe cancellation in the observables due to the zero-temperature subtraction procedure at large $N_t$. 
Nevertheless, we think it is worth taking advantage of smaller overall simulation costs with the fixed-scale approach to calculate the EOS in 2+1 flavor QCD with small discretization errors around $T_{\rm pc}$.

Another point of our study is the choice of the quark action on the lattice.
Most lattice studies of hot/dense QCD have been done with computationally 
less expensive staggered-type lattice quarks \cite{Borsanyi:2010cj,Bazavov:2010sb}. 
However, their theoretical basis such as locality and universality are not well established. 
Therefore, to check the validity of these results
it is important to compare the results with those obtained using theoretically sound lattice quarks, such as the Wilson-type quarks.
See \cite{DIK2010,Conf10,tmfT11,BWLat11} for recent studies of QCD thermodynamics with Wilson-type quarks.
A systematic study of the EOS with Wilson-type quarks has been done so far only in the case of two-flavor QCD \cite{AliKhan:2000,AliKhan:2001ek}.
We extend the study to the more realistic case of 2+1 flavor QCD, using a nonperturbatively improved Wilson quark action coupled to a RG-improved Iwasaki gauge action.

Thanks to the fixed-scale approach, we can take advantage of using the zero-temperature configurations on the ILDG. 
Using the same combination of lattice actions as ours, the CP-PACS+JLQCD Collaboration has generated a set of zero-temperature configurations in 2+1 flavor QCD and has studied their hadronic spectrum \cite{Aoki:2005et,Ishikawa:2007nn}. 
Another attractive point of the fixed-scale approach in a study with improved Wilson quarks is that, unlike the case of the fixed-$N_t$ approach, we can keep the lattice spacing small at all temperatures and thus can avoid extrapolating the nonperturbative clover coefficient $c_{\rm SW}$ to coarse lattices on which the improvement program is not quite justified.

Choosing a simulation point of the CP-PACS+JLQCD Collaboration, we carry out finite-temperature simulations to perform the first calculation of the EOS in 2+1 flavor QCD with improved Wilson quarks.
Although the light quark masses studied are still heavier than their physical values, 
we find that the EOS obtained is roughly consistent with recent results using highly improved staggered 
quarks in the fixed-$N_t$ approach at large values of $N_t$.

In  the next section, we introduce the $T$-integration method which enables us to calculate the EOS nonperturbatively in the fixed-scale approach.
The lattice setup and the simulation parameters are summarized in Sec.~\ref{sec:setup}.
Results of gauge observables are presented in Sec.~\ref{sec:gauge}.
In Sec.~\ref{sect:beta} the beta functions are evaluated.
Our results on the EOS are shown in Sec.~\ref{sect:eos} and a summary is given in Sec.~\ref{sec:summary}.
The Appendix~\ref{app1} is devoted to a discussion about the choice of the interpolation procedure for the $T$-integration method. 
Preliminary results of this study have been reported in \cite{Kanaya:2009nf,Umeda:2010ye}.

\section{$T$-integration method}
\label{sec:Tint}

In conventional studies of EOS in the fixed-$N_t$ approach, the pressure $p$ is nonperturbatively estimated by the ``integration method'' \cite{Engels:1990vr}:
\begin{eqnarray}
p = \frac{T}{V} \int^{\vec{b}}_{\vec{b}_0} \! d\vec{b} \cdot \left\langle \frac{1}{Z}
\frac{\partial Z}{\partial \vec{b} } \right\rangle_{\rm \! sub}
= -\frac{T}{V} \int^{\vec{b}}_{\vec{b}_0} \! d\vec{b} \cdot \left\langle
\frac{\partial S}{\partial \vec{b}} \right\rangle_{\rm \! sub}
\label{eq:Integral}
\end{eqnarray}
where $V$ is the spatial volume of the system, $Z$ is the partition function, $S$ is the lattice action with the coupling parameters $\vec{b} = (\beta, \kappa_{ud}, \kappa_s, \cdots)$, and $\langle \cdots \rangle_{\rm sub}$ is the thermal average with a zero-temperature contribution subtracted for renormalization. 
This relation is obtained by differentiating and then integrating the thermodynamic relation $p = (T/V) \ln Z$ in the coupling parameter space of $\vec{b}$. 
The initial point $\vec{b}_0$ is chosen in the low temperature phase such that $p(\vec{b}_0) \approx 0$.

This method is inapplicable in the fixed-scale approach because $\vec{b}$ is fixed in the simulations.
To overcome the problem, we have developed the ``$T$-integration method''  \cite{Umeda:2008bd}:
Using a thermodynamic relation at vanishing chemical potential,
\begin{eqnarray}
T \frac{\partial}{\partial T} \left( \frac{p}{T^4} \right) =
\frac{\epsilon-3p}{T^4},
\end{eqnarray}
where $\epsilon$ is the energy density, we obtain another nonperturbative estimate of the pressure, 
\begin{eqnarray}
\frac{p}{T^4} = \int^{T}_{T_0} dT \, \frac{\epsilon - 3p}{T^5},
\label{eq:Tintegral}
\end{eqnarray}
with the initial temperature $T_0$ chosen such that $p(T_0) \approx 0$.
Here, the trace anomaly $\epsilon -3p$ is calculated as
\begin{equation}
\frac{\epsilon-3p}{T^4}=
\frac{1}{T^3 V} \; a\frac{d \vec{b}}{d a} \cdot
\left\langle \frac{\partial S}{\partial\vec{b}} \right\rangle_{\rm \! sub}
\label{eq:e-3p:general}
\end{equation}
where $a(d\vec{b}/da)$ is a vector of the beta functions which describes the variation of $\vec{b}$ along the LCP.

When we vary $T$ along a LCP by varying $\vec{b}$, the integral in (\ref{eq:Tintegral}) is equivalent to that in (\ref{eq:Integral}), with the integration path chosen to be on the same LCP.
However, (\ref{eq:Tintegral}) allows us to vary $T$ without varying $\vec{b}$.
In the fixed-scale approach, we vary $T$ by varying $N_t$.
Because $N_t$ is discrete, we have to interpolate the data with respect to $T$ to carry out the integration of (\ref{eq:Tintegral}).
The systematic error from the interpolation should be checked.

In \cite{Umeda:2008bd}, the $T$-integration method was tested in quenched QCD 
and it was shown that the systematic error from the discreteness of $T$ is under control 
when $a$ is chosen sufficiently small, as adopted in spectrum studies.
The EOS from the fixed-scale approach was shown to be well consistent with that from the fixed-$N_t$ approach with large $N_t$ ($N_t \ge 8$), except for the high temperature limit where the fixed-scale approach suffers from lattice discretization errors, 
as discussed in Sec.~\ref{sect1}.

\section{Lattice setup}
\label{sec:setup}

\begin{figure}[tb]
\includegraphics[width=110mm]{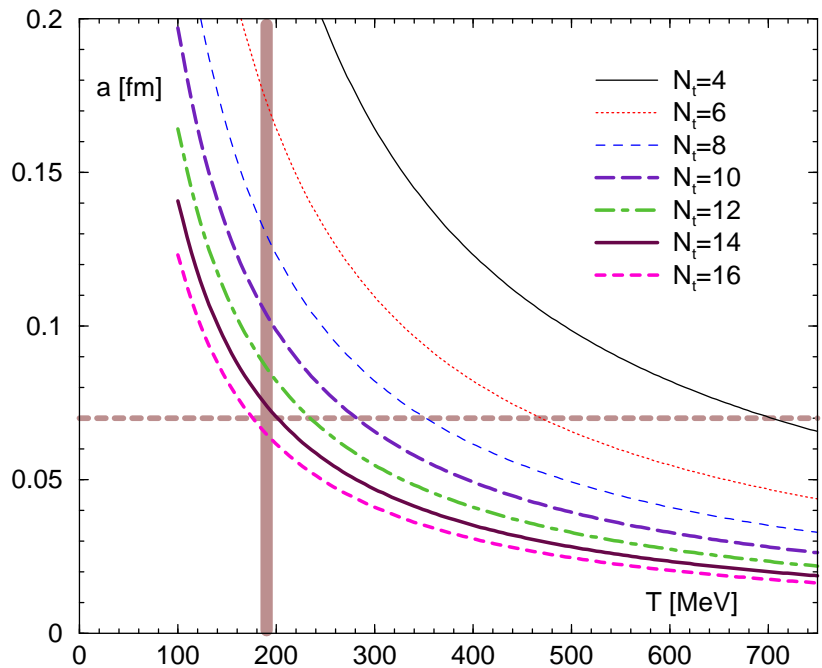}
\caption{Temperature vs lattice spacing at each $N_t$. 
The horizontal dashed line at $a \simeq 0.07$ fm represents the lattice spacing in this study.
The vertical shaded line represents the approximate location of the pseudo-critical temperature at our quark masses. 
}
\label{fig:t_vs_a}
\end{figure}

\begin{figure}[tb]
\includegraphics[width=110mm]{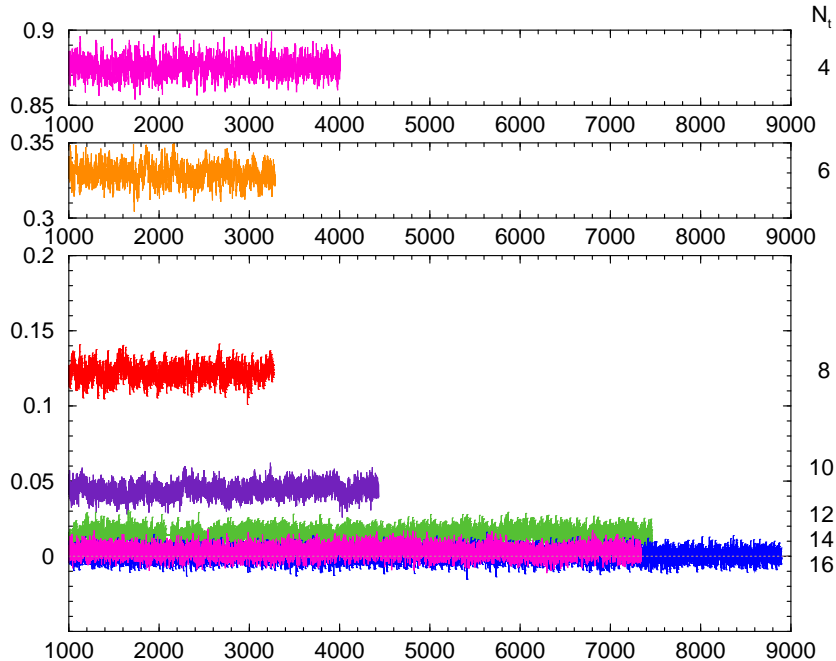}
\caption{Time history of the Polyakov loop measured on finite-temperature 
lattices. The horizontal axis is the trajectory length.
}
\label{fig:history}
\end{figure}

\begin{table*}[tb]
\begin{tabular}{ccccccllll}
\hline
$N_t$ & $T$[MeV] & $\delta\tau$ & Trajectory & Thermalization & Bin size & \hspace{0mm}Plaquette & \hspace{0mm}Rectangular
& \hspace{5mm}$\langle L \rangle$ & \hspace{5mm}$\chi_L$
\\
\hline
58& --- & --- & 6500 & ---                   &  ---     & 0.6040260( 40) &
0.3770800( 50)  & \hspace{6mm}--- & \hspace{5mm}---\\
\hline
16& 174 & 1/140 &  7895 & 1000  & 500 & 0.6040337( 50) &
0.3770870(86)  & 0.000213(21) &  0.0018(20) \\
14& 199 & 1/120 &  6370 & 1000  & 500 & 0.6041040(100) &
0.3772003(168) & 0.001172(67)   &  0.0075(34)  \\
12& 232 & 1/120 &  6460 & 1000  & 300 & 0.6041789( 53) &
0.3773145(80)  & 0.004911(60)   &  0.0141(25)  \\
10& 279 & 1/90 &  3935 &   500    & 200 & 0.6042629( 50) &
0.3774460(86) & 0.01470(11)   &  0.0528(58)  \\
8& 348 & 1/60 &  2770 &   500      & 100 & 0.6043430( 87) &
0.3775803(141) & 0.04072(12)   &  0.115(13)  \\
6& 464 & 1/52 &  2785 &   500      &   50 & 0.6045902( 93) &
0.3780182(150) & 0.10981(11)   &  0.190(15) \\
4& 697 & 1/44 &  3510 &   500      &   50 & 0.6061122( 93) &
0.3809620(144) & 0.291854(74)   &  0.2168(92) \\
\hline
\end{tabular}
\caption{Simulation parameters and gauge observables. 
The zero-temperature results ($N_t=58$) are taken from \cite{Ishikawa:2007nn} by the CP-PACS+JLQCD Collaboration.
Temperature $T$ is determined using $1/a=2.79$ GeV ($a \simeq 0.07$fm) \cite{Ishikawa:2007nn}.
The Metropolis test is performed every 0.5 trajectories for 
finite-temperature simulations.
$\delta \tau$ is the molecular dynamics time step, and 
``Bin size'' is the bin size for gauge observables, both in units of 
trajectories. ``Trajectory'' is the generated trajectory length after 
thermalization of  ``Thermalization'' trajectories.
``Plaquette'' and ``Rectangular'' are plaquette and rectangular loop expectation values.
$\langle L \rangle$ and $\chi_L$ are the bare Polyakov loop and its susceptibility, respectively.
}
\label{tab1}
\end{table*}

We adopt a nonperturbatively $O(a)$-improved Wilson quark action \cite{Sheikholeslami:1985ij} coupled with the RG-improved Iwasaki gauge action \cite{Iwasaki} to simulate 2+1 flavor QCD:
\begin{eqnarray}
S_g &=& -\beta \sum_{x} \left\{ 
\sum_{\mu>\nu}c_0W^{1\times 1}_{\mu\nu}(x)
+\sum_{\mu,\nu}c_1W^{1\times 2}_{\mu\nu}(x)
\right\},\label{eq:gaction}\\
S_q &=& \sum_{f=u,d,s}\sum_{x,y} \bar{q}_x^f D_{xy}^f q_y^f, \label{eq:qaction}\\
D_{xy}^f &=& \delta_{x,y}-\kappa_f
\sum_\mu\{ (1-\gamma_\mu)\,U_{x,\mu}\delta_{x+\hat{\mu},y} + 
(1+\gamma_\mu)\,U^\dagger_{x-\hat{\mu},\mu}\delta_{x-\hat{\mu},y}
\}
-\delta_{x,y}\, c_{\rm SW}(\beta)\, \kappa_f\sum_{\mu>\nu}\sigma_{\mu\nu}
F_{\mu\nu}
\end{eqnarray}
with $\kappa_u = \kappa_d \equiv \kappa_{ud}$.
The clover coefficient $c_{\rm SW}(\beta)$ has been evaluated nonperturbatively by the Schr\"{o}dinger functional method in \cite{Aoki:2005et}. 
Hadronic properties have been systematically studied  with this action by the CP-PACS, JLQCD and PACS-CS Collaborations, down to the physical point \cite{Ishikawa:2007nn, Aoki:2009sf, Aoki:2009pp, Ishikawa:2009su, Aoki:2009ix}.

In this study, we use the zero-temperature configurations by the CP-PACS and JLQCD Collaborations \cite{Ishikawa:2007nn}, 
which are open to the public at ILDG/JLDG \cite{Maynard:2010wi}.
The CP-PACS+JLQCD zero-temperature configurations are available at three $\beta$'s, five $\kappa_{ud}$'s, and two $\kappa_s$'s, i.e. at a total of 30 simulation points.
Among them, we choose $\beta=2.05$, $\kappa_{ud}=0.1356$, and $\kappa_s=0.1351$, 
which correspond to the smallest lattice spacing and the lightest $u$ and $d$ quark masses ($m_\pi/m_\rho\simeq0.63$) with $m_s$ near its physical point ($m_{\eta_{ss}}/m_{\phi}\simeq0.74$). 
The hadronic radius is estimated to be $r_0/a=7.06(3)$ \cite{Maezawa:2009di}.
Setting the lattice scale by $r_0=0.5$ fm, we estimate the scale as $1/a\simeq 2.79$ GeV ($a \simeq 0.07$fm).
The lattice size is $28^3 \times 56$ ($N_s a \simeq 2$ fm), and the number of thermalized configurations are 650 (6500 trajectories), which are stored  every 10 trajectories.
Note that the $u$ and $d$ quark masses are still much larger than their physical values.
We are planning to extend the study down to the physical point \cite{Aoki:2009ix}.

Adopting the same coupling parameters as the zero-temperature simulation \cite{Ishikawa:2007nn}, 
we generate finite-temperature configurations on 
$32^3\times N_t$ lattices with $N_t=4$, 6, $\cdots$, 16. 
Our generation code is based on the Colombia Physics System (CPS) code \cite{CPS} with the RHMC algorithm for the $s$ quark. 
We tuned the acceptance rate at the Metropolis test to be about 80\%.
The simulation parameters are summarized in Table~\ref{tab1}. 

Using the relation between $T$ and $N_t$, our range of $N_t$ corresponds to the range $T=174$--697 MeV at $\beta=2.05$, as shown in Fig.~\ref{fig:t_vs_a}.
Previous studies of the pseudocritical temperature $T_{\rm pc}$ in two-flavor QCD with improved Wilson quarks at $N_t \sim 6$ \cite{Ejiri:2009hq,DIK2010} suggest $T_{\rm pc}$ around 200 MeV for $m_\pi/m_\rho\simeq0.63$ in two-flavor QCD.
Taking into account the effect of the dynamical $s$ quark and also our larger values of $N_t\sim14$ around the pseudocritical point, we expect a smaller value for $T_{\rm pc}$. 
In the succeeding sections, we show that our data suggest $T_{\rm pc}\sim190$ MeV at our simulation point, as shown in Fig.~\ref{fig:t_vs_a} by the vertical shaded line.

The fixed-scale approach is not applicable at very high temperatures, where the lattice spacing $a$ becomes too coarse to resolve thermal fluctuations \cite{Umeda:2008bd}.
We may estimate a typical length scale of thermal fluctuations by the thermal wave length $\lambda \sim 1/ E$, where $E$ is an average energy of massless particles at finite $T$. 
We then obtain $\lambda \sim 1/(3T)$ 
from $E \sim 3 T \zeta(4)/\zeta(3) \sim 2.7 T$ for the Bose-Einstein distribution and $E \sim 3 T \zeta(4)/\zeta(3) \times 7/6 \sim 3.15 T$ for the Fermi-Dirac distribution.
Thus, data at $T \simge 1/(3a)$ should be taken with care \cite{Maezawa2011}.
On the present lattice, the data at $T\simeq700$ MeV may suffer from some lattice artifacts.

\section{Gauge observables}
\label{sec:gauge}

\begin{figure}[tb]
\includegraphics[width=85mm]{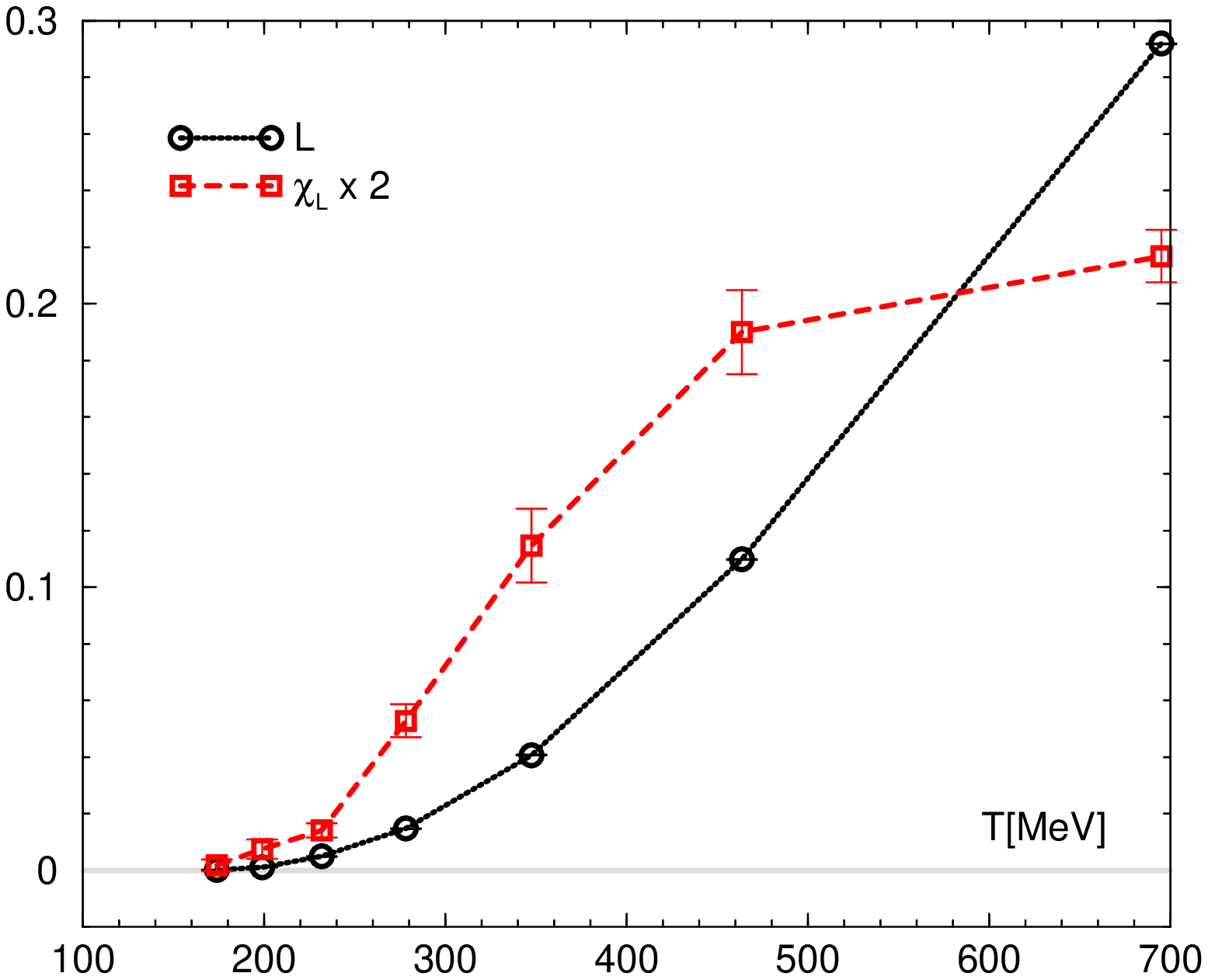}
    \hspace{3mm}
\includegraphics[width=85mm]{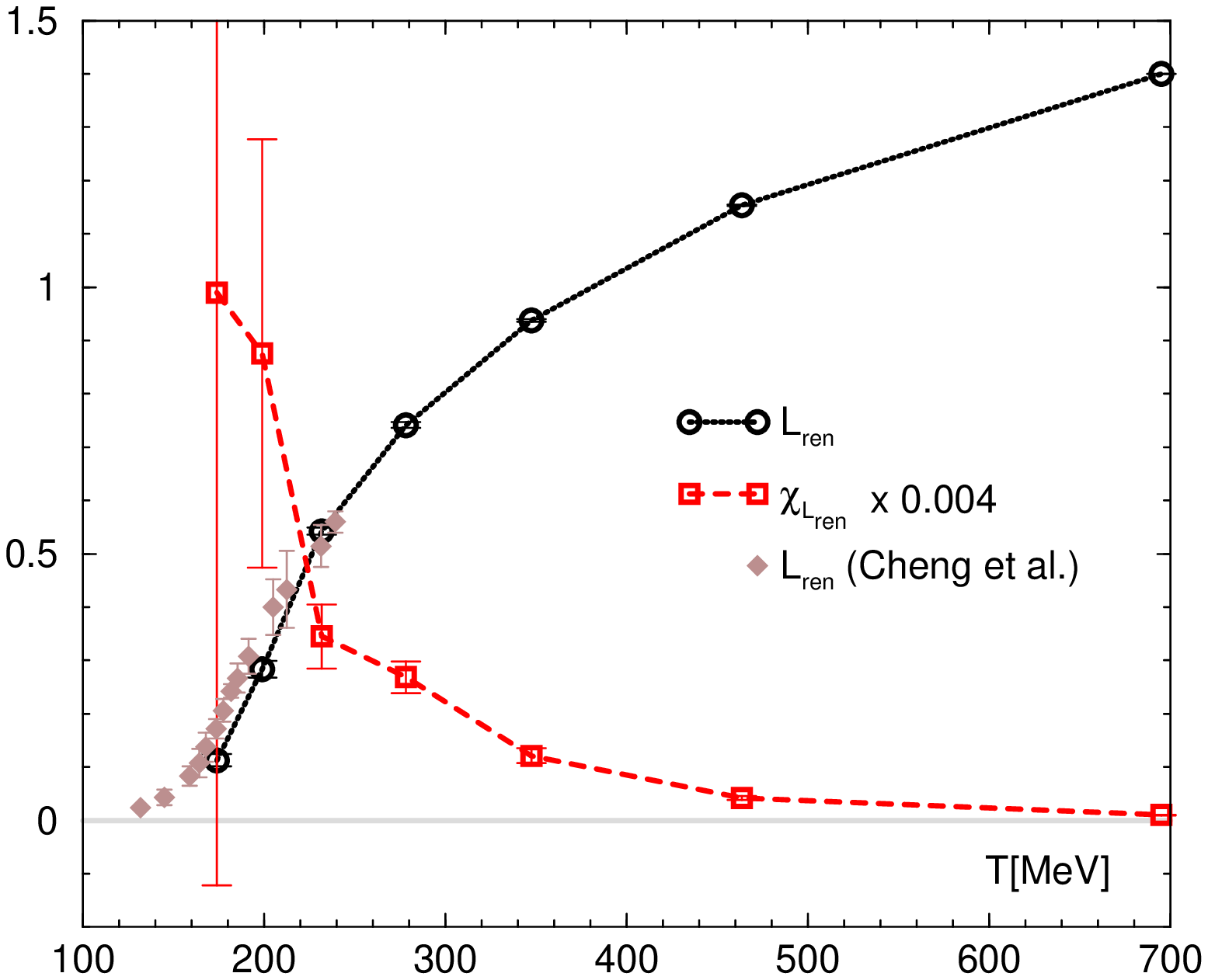}
\caption{Polyakov loop expectation value and its susceptibility as functions of $T$. 
The left panel shows the bare results; the right panel shows renormalized results using the renormalization scheme of  \cite{RBCB2008}.
$\chi_{L}$ is multiplied by $2$ and 
$\chi_{L_{\rm ren}}$ is multiplied by $0.004$ to fit into the same scale.
Also shown in the right panel are the results of $\langle L_{\rm ren}\rangle$ from the p4 staggered quark action obtained at $m_{ud}^{\rm bare}/m_s^{\rm bare} = 0.05$ in the fixed $N_t$ approach at $N_t=8$ \cite{RBCB2010}, where the horizontal axis is rescaled using $r_0=0.5$ fm.}
\label{fig:poly_sus}
\end{figure}

The expectation values of gauge observables are measured every 0.5 trajectories.
The results of basic observables are summarized in Table~\ref{tab1}. 
The time history of the Polyakov loop defined by  
\begin{equation}
L=\frac{1}{V}\sum_{\vec{x}} \frac{1}{3}{\rm Tr} \prod_{\tau=1}^{N_t}U_{(\tau,\vec{x}),4}
\end{equation}
is shown in Fig.~\ref{fig:history}.
The gauge configurations are stored every five trajectories, on which quark observables are measured. 
By examining the bin-size dependence of the errors,  we estimate the statistical errors for gauge observables by the jackknife method with the bin size listed in Table~\ref{tab1},
while those for quark observables are estimated with the bin size of 25 trajectories after thermalization of 1000 trajectories.
Static quark potentials measured on the same configurations are studied in \cite{Maezawa:2009di,Maezawa2011}.
In the following, we disregard the statistical error in $T$ from that of the lattice scale $a$, which is about 0.5\%.
Note that, because the scale is common for all $T$'s in the 
fixed-scale approach, a shift in the scale $a$ just causes an overall shift of $T$.

The left panel of Fig.~\ref{fig:poly_sus} shows the results of the Polyakov loop 
expectation value $\langle L \rangle$ and its susceptibility $\chi_L=N_s^3 (\langle L^2 \rangle-\langle L \rangle^2)$
as functions of $T$.
We find that $\langle L \rangle$ starts deviating from zero at $T\sim180$--200 MeV, suggesting the pseudocritical point around there.

For a comparison with the results of previous studies in the fixed-$N_t$ approach, we have to renormalize $\langle L \rangle$.
Although the additive renormalization constant for free energies is independent of $T$ and thus is common for all $T$'s in the fixed-scale approach, the Polyakov loop $\langle L \rangle \sim e^{-F/T}$ does receive a $T$-dependent renormalization. 
To enable a direct comparison with the results of staggered-type quarks, we adopt the renormalization scheme proposed in \cite{RBCB2008}; {\it i.e.} we renormalize $L$ such that the singlet free energy from $L_{\rm ren} = (Z_{\rm ren})^{N_t} L$ becomes the L\"uscher's universal bosonic-string potential $ - \pi/(12 r) + \sigma r$ at $r=1.5\,r_0$ \cite{Luescher}, where $\sigma$ is the string tension at $T=0$.
Using our potential data at $T=0$ \cite{Maezawa:2009di}, we obtain $Z_{\rm ren}=1.4801(90)$.
Our results for $\langle L_{\rm ren} \rangle$ and the corresponding susceptibility $\chi_{L_{\rm ren}}$ are plotted in the right panel of Fig.~\ref{fig:poly_sus}. 
We note that the dependences on $T$ in these quantities are largely influenced by the renormalization factor.
In spite of the heavier light quark mass in our study, our results for $\langle L_{\rm ren} \rangle$ agree well with a result from the p4 staggered quark action in the fixed-$N_t$ approach at $N_t = 8$ \cite{RBCB2010} (see the right panel of Fig.~\ref{fig:poly_sus}).
Similar agreement of $\langle L_{\rm ren} \rangle$ between a smeared  Wilson-type quark action and a smeared staggered-type quark action is reported in \cite{BWLat11}.

In Fig.~\ref{fig:poly_sus}, we also show the results of Polyakov loop 
susceptibilities. In the left panel of Fig.~\ref{fig:poly_sus}, besides a 
faint bump at $T\sim200$ MeV, we do not see a clear signal of a peak in 
$\chi_L$ at the two discrete simulation points in the range 180-200 MeV 
where $T_{\rm pc}$ is expected.
In the right panel of Fig.~\ref{fig:poly_sus}, existence of a peak of $\chi_{L_{\rm ren}}$ 
around these temperatures is not excluded, but due to the large errors there.
The origin of the large errors will be discussed in Sect.~\ref{sect:eos} 
and \ref{sec:summary}.
This is in contrast with the case of our previous study in quenched QCD adopting the fixed-scale approach \cite{Umeda:2008kz}, in which we observe a clear peak of $\chi_L$, and also with the cases of full QCD studies adopting the fixed-$N_t$ approach with staggered-type (see e.g.\ \cite{RBCBi2007}) and Wilson-type \cite{Ejiri:2009hq,DIK2010} quarks.
As a possible cause of the absence of a clear peak in this study, we note that the resolution in $T$ is lower than that in our previous quenched study.
We may have missed the peak between the simulation points.
We also note the following:
(i) We probably have a crossover in full QCD around the simulated quark masses instead of the first-order deconfining transition in quenched QCD.  
(ii) Our previous experience with improved Wilson quarks suggests that the peak becomes milder with increasing $N_t$. Our $N_t \sim 14$ around the crossover point is larger than those adopted in previous studies with the fixed-$N_t$ approach.
(iii) The aspect ratio $N_s/N_t$ is not large at low temperatures in this study.
All of these will make the peak milder and thus more difficult to detect when the resolution in $T$ is not fine enough.

\section{Beta functions}
\label{sect:beta}

\begin{figure}[bt]
  \begin{center}
    \includegraphics[width=110mm]{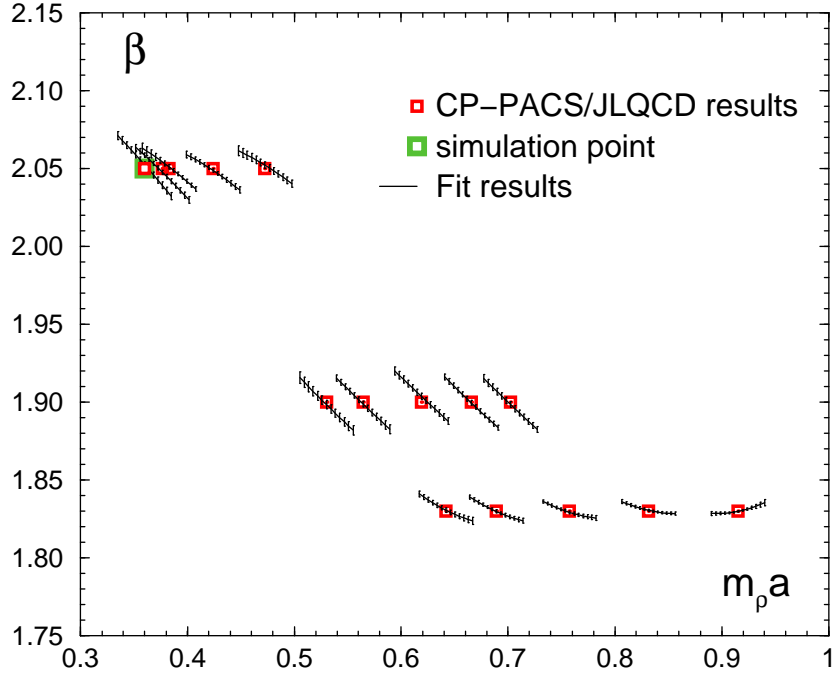}
    \caption{
    The global fit for coupling parameters, $\beta$, 
    as a function of $m_\rho a$. Square symbols show coupling 
    parameters in the CP-PACS/JLQCD study. The solid lines show the global
    fit results for each simulation point with corresponding 
    $m_\rho/m_\pi$ and $m_{\eta_{ss}}/m_\phi$. 
    To avoid a plot that is too busy, only half of the data points are shown ($\kappa_s=0.1371$, 0.1358, and 0.1351 at $\beta=1.83$, 1.90, and 2.05, respectively).
    }
    \label{fig:fit1}
  \end{center}
\end{figure}

\begin{figure}[bt]
  \begin{center}
    \includegraphics[width=85mm]{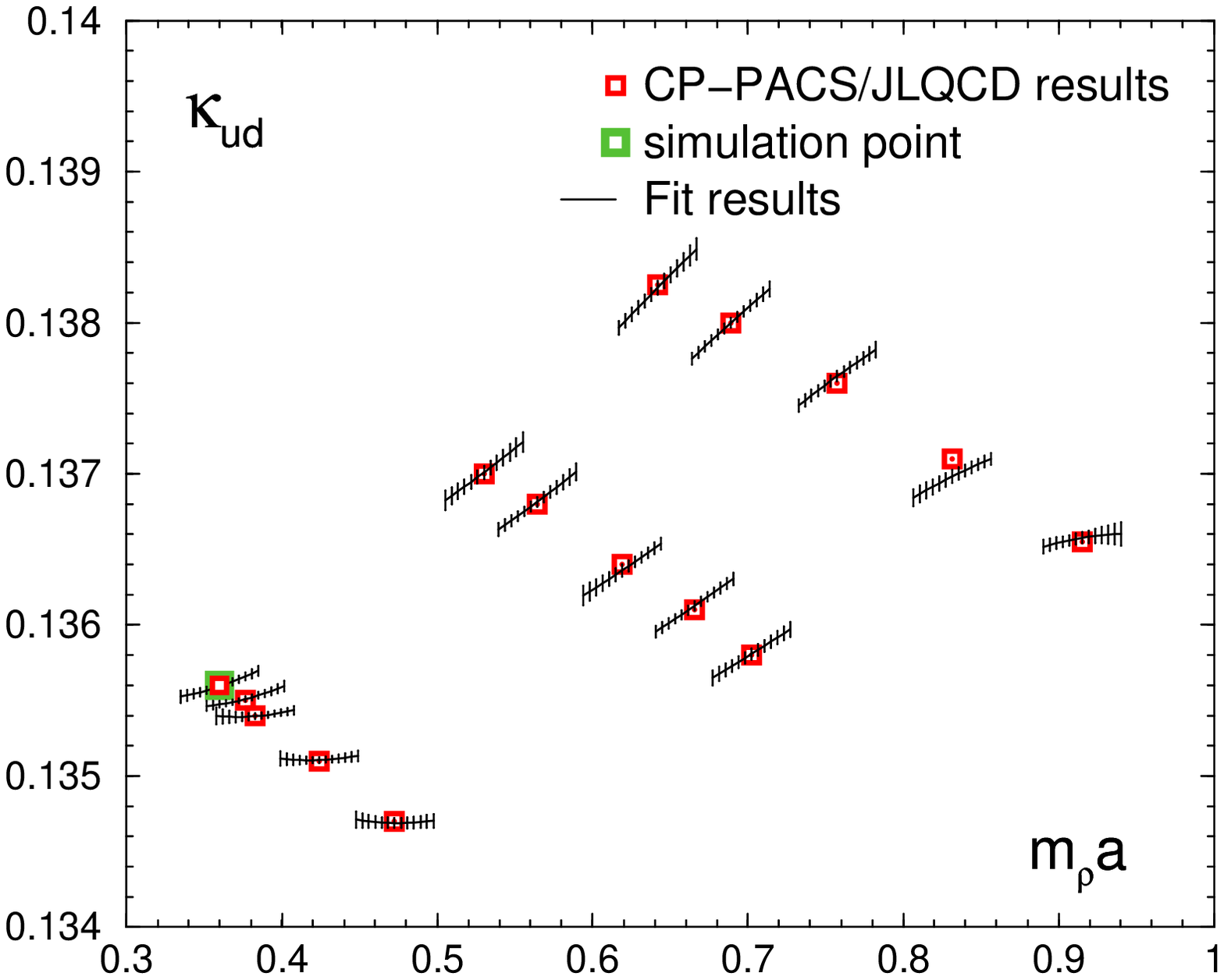}
    \hspace{3mm}
    \includegraphics[width=85mm]{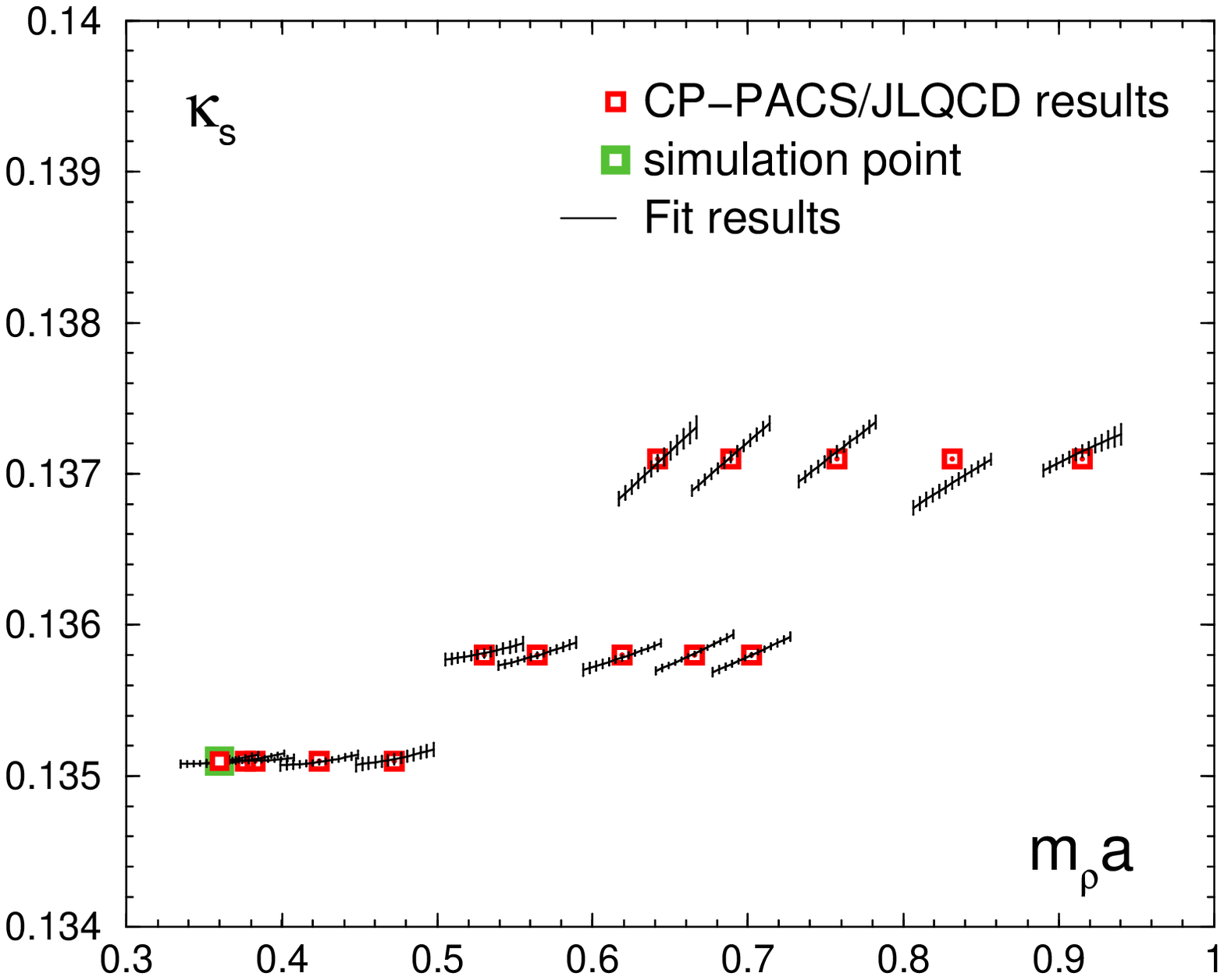}
    \caption{The same as Fig.~\ref{fig:fit1} but for $\kappa_{ud}$ and  $\kappa_{s}$}
    \label{fig:fit2}
  \end{center}
\end{figure}

\begin{figure}[bt]
  \begin{center}
    \includegraphics[width=110mm]{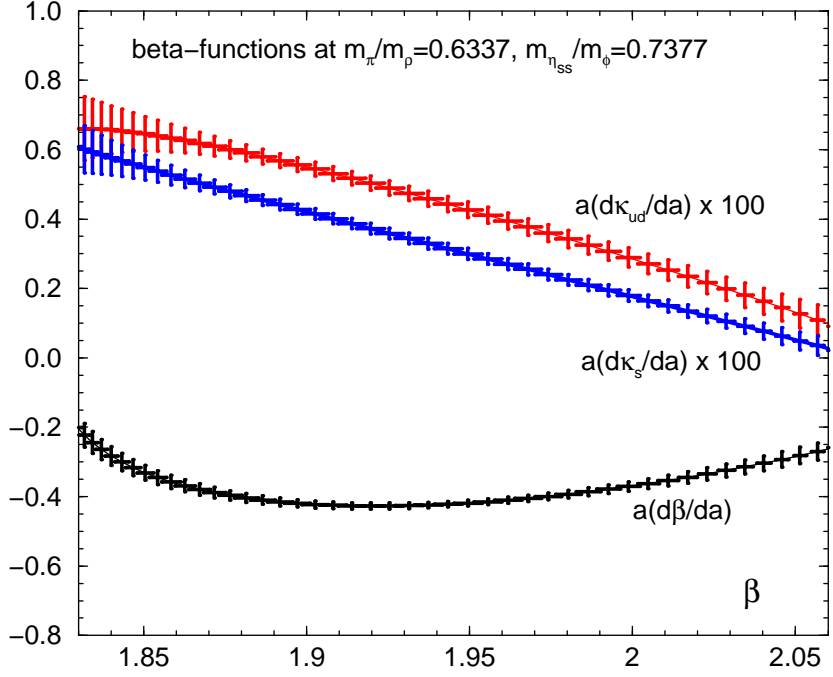}
    \caption{Beta functions on our LCP, $m_\pi/m_\rho=0.6337$ and $m_{\eta_{ss}}/m_{\phi}=0.7377$, as functions of $\beta$. The scale setting is made with $am_\rho$. 
    Beta functions for $\kappa_{ud}$ and $\kappa_s$ are magnified by a factor of 100.
    Horizontal and vertical bars at each data point represent statistical errors.}
    \label{fig:bfunc2}
  \end{center}
\end{figure}

\begin{figure}[bt]
  \begin{center}
    \includegraphics[width=110mm]{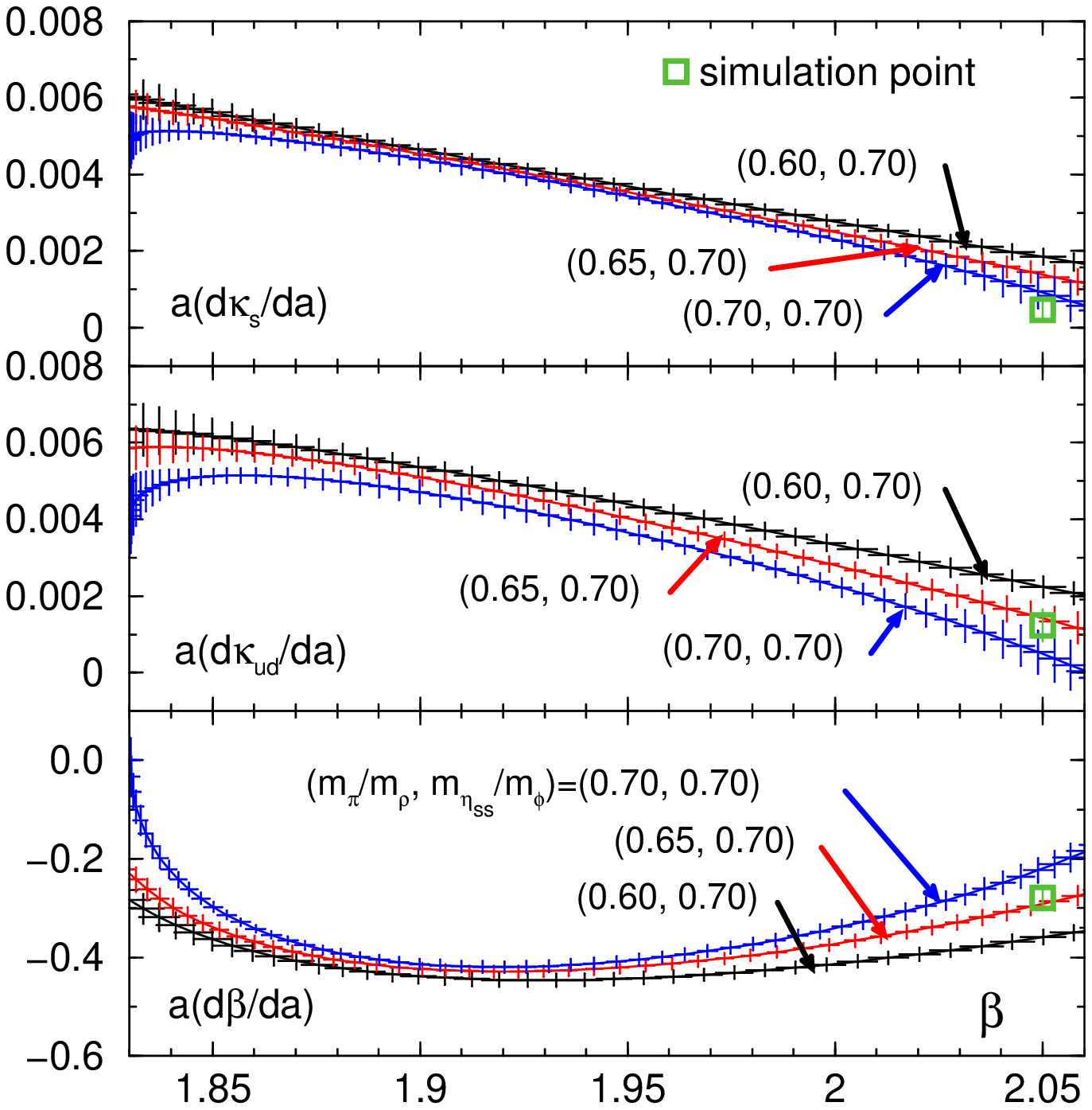}
    \caption{Beta functions at light quark masses corresponding to $m_\pi/m_\rho\simeq0.5$, 0.6, and 0.7, with the $s$ quark mass $m_{\eta_{ss}}/m_{\phi}\simeq0.7$, as functions of $\beta$.
    The simulation point of this study is marked by an open square.
    Horizontal and vertical bars at each data point represent statistical errors.}
    \label{fig:bfunc3}
  \end{center}
\end{figure}

To evaluate the trace anomaly according to (\ref{eq:e-3p:general}), we need the beta functions $a(d\beta/da)$ and $a(d\kappa_f/da)$ ($f=ud$ and $s$).
In this study, we define LCP's by $m_\pi/m_\rho$ and $m_{\eta_{ss}}/m_\phi$ at $T=0$.
The beta functions are determined nonperturbatively through the coupling parameter dependence of zero-temperature observables.
We use the data of $am_\rho$, $m_\pi/m_\rho$, and $m_{\eta_{ss}}/m_\phi$ at 30 simulation points of the CP-PACS+JLQCD zero-temperature configurations \cite{Ishikawa:2007nn} to extract the beta functions.
From a previous experience of two-flavor QCD with improved Wilson quarks in the fixed-$N_t$ approach \cite{AliKhan:2001ek}, we expect that, although $a(d\kappa_f/da)$'s are much smaller than $a(d\beta/da)$, in the trace anomaly, 
the overall magnitude of the quark contribution proportional to $a(d\kappa_f/da)$ is comparable with that of the gauge part proportional to $a(d\beta/da)$, but with opposite sign.
Therefore, evaluation of the quark contribution is important.

In our previous attempt \cite{Kanaya:2009nf}, we have tried to evaluate the beta functions by the inverse matrix method, which was successful in the case of two-flavor QCD \cite{AliKhan:2001ek}.
In 2+1 flavor QCD, we fitted the data of $am_\rho$, $m_\pi/m_\rho$, and $m_{\eta_{ss}}/m_\phi$ 
as functions of three coupling parameters ($\beta$, $\kappa_{ud}$, $\kappa_s$), and 
inverted the matrix of the slopes of the former in terms of the latter to obtain the beta functions.
However, it turned out that errors in $a(d\kappa_f/da)$ are too large to calculate the quark part EOS reliably, although the magnitude of the beta functions and the result for the gauge part of the trace anomaly are consistent with an expectation from the two-flavor case \cite{Kanaya:2009nf}. 
The situation is also similar when we use the data of pseudoscalar decay constants instead of $m_{\rho}$.
We find that the large errors in $a(d\kappa_f/da)$ are mainly due to the matrix inversion procedure, through which all components of the inverse matrix get errors of similar magnitude. 
Because $a(d\kappa_f/da)$ are much smaller than $a(d\beta/da)$, we need more precise values of the slopes to suppress the errors in $a(d\kappa_f/da)$.
In the present case of 2+1 flavor QCD, the data points of zero-temperature configurations around the simulation point are not dense enough to achieve the required precision of the slopes. 

To avoid the matrix inversion procedure, we now adopt an alternative method, the direct fit method \cite{AliKhan:2001ek}: 
We fit the coupling parameters, $\beta$, $\kappa_{ud}$, and $\kappa_s$, 
as a function of three observables, $am_\rho$, $m_\pi/m_\rho$, and $m_{\eta_{ss}}/m_\phi$. 
Consulting the overall quality of the fits, we adopt the following third order polynomial function of the observables in this study: 
\begin{eqnarray}
\left(
\begin{array}{c}
\beta \\
\kappa_{ud} \\
\kappa_s
\end{array}
\right)
&=& \vec{c}_0 + \vec{c}_1\,(am_\rho) + \vec{c}_2\,(am_\rho)^2
+ \vec{c}_3\left(\frac{m_\pi}{m_\rho}\right)
+ \vec{c}_4\left(\frac{m_\pi}{m_\rho}\right)^2
+ \vec{c}_5 \,(am_\rho)\left(\frac{m_\pi}{m_\rho}\right)
\nonumber\\
&&+\;  \vec{c}_6\left(\frac{m_{\eta_{ss}}}{m_\phi}\right)
+ \vec{c}_7\left(\frac{m_{\eta_{ss}}}{m_\phi}\right)^2
+ \vec{c}_8\,(am_\rho)\left(\frac{m_{\eta_{ss}}}{m_\phi}\right)
+ \vec{c}_9\left(\frac{m_\pi}{m_\rho}\right) \left(\frac{m_{\eta_{ss}}}{m_\phi}\right)
\nonumber\\
&&+\; \vec{c}_{10}\,(am_\rho)^3
+ \vec{c}_{11}\left(\frac{m_\pi}{m_\rho}\right)^3
+ \vec{c}_{12}\left(\frac{m_{\eta_{ss}}}{m_\phi}\right)^3
+\; \vec{c}_{13} \,(am_\rho)\left(\frac{m_\pi}{m_\rho}\right)^2
\nonumber\\
&&+\; \vec{c}_{14}\,(am_\rho)^2\left(\frac{m_\pi}{m_\rho}\right)
+ \vec{c}_{15}\,(am_\rho)\left(\frac{m_{\eta_{ss}}}{m_\phi}\right)^2
+ \vec{c}_{16}\,(am_\rho)^2\left(\frac{m_{\eta_{ss}}}{m_\phi}\right) 
\nonumber\\
&&+\; \vec{c}_{17}\left(\frac{m_\pi}{m_\rho}\right) \left(\frac{m_{\eta_{ss}}}{m_\phi}\right)^2
+ \vec{c}_{18}\left(\frac{m_\pi}{m_\rho}\right)^2
\left(\frac{m_{\eta_{ss}}}{m_\phi}\right)
+ \vec{c}_{19}\,(am_\rho)\left(\frac{m_\pi}{m_\rho}\right)
\left(\frac{m_{\eta_{ss}}}{m_\phi}\right).
\label{eq:beta}
\end{eqnarray}
Note that the fits for the three coupling parameters are independent of each other.
Figures \ref{fig:fit1} and \ref{fig:fit2} show the results of the global 
fit (\ref{eq:beta}) as functions of $m_\rho a$.  
The fits with dof $=10$ lead to reasonable $\chi^2/$dof
($=1.63$, 1.08, and 1.69 for the fit of $\beta$, $\kappa_{ud}$, and
$\kappa_s$, respectively), where the standard deviation of each coupling
parameter is estimated by the error propagation rule using the errors
of the observables and the partial derivatives of the resulting fitting
function, Eq.(9), with respect to the observables, neglecting the covariance
among the observables.

\begin{table*}[tb]
\begin{tabular}{c|cccccc}
\hline
scale setting & $\displaystyle{ a\frac{d \beta}{d a} }$ & $\chi^2/$dof & $\displaystyle{a\frac{d \kappa_{ud}}{da} }$ & $\chi^2/$dof & $\displaystyle{ a\frac{d\kappa_s \strut}{da \strut} }$ & $\chi^2/$dof 
\\
\hline
$am_\rho$ & -0.279(24) & 1.6 & 0.00123(41) & 1.1 & 0.00046(26) & 1.7 \\
$am_\pi$ & -0.319(21) & 1.2 & 0.00179(38) & 0.8 & 0.00088(22) & 1.3 \\
$am_{K}$ & -0.252(25) & 1.0 & 0.00105(44) & 1.0 & 0.00043(32) & 1.3 \\
$am_{K^*}$ & -0.215(28) & 1.1 & 0.00055(47) & 1.2 & 0.00002(36) & 1.8 \\
\hline
\end{tabular}
\caption{Beta functions at our simulation point determined by the global fit (\ref{eq:beta}) or with alternative scale setting variables. Values of $\chi^2/$dof for the fits are also given.}
\label{tab:betafunc}
\end{table*}

We define the LCP by fixing $m_\pi/m_\rho$ and $m_{\eta_{ss}}/m_\phi$.
Then, the beta functions are calculated as
$a\,d \beta/d a = (a m_\rho)\, \partial \beta/\partial (a m_\rho)$, etc., in terms of the coefficients $\vec{c}_1$, $\vec{c}_2$, $\vec{c}_5$, $\vec{c}_8$, $\vec{c}_{10}$, etc.\ in (\ref{eq:beta}).
The resulting beta functions for our LCP  ($m_\pi/m_\rho=0.6337$, $m_{\eta_{ss}}/m_{\phi}=0.7377$) are shown in Fig.~\ref{fig:bfunc2} as functions of $\beta$.
Beta functions at other light quark masses are shown in Fig.~\ref{fig:bfunc3}. 
As the variable to set the scale, we may alternatively adopt $am_\pi$, $am_{K}$, or $am_{K^*}$ instead of $am_\rho$ in (\ref{eq:beta}).
Results of the beta functions, at our simulation point ($\beta=2.05$ on our LCP), adopting various scale setting variables are listed in Table~\ref{tab:betafunc}.
Taking the results from $am_\rho$ as the central value, we obtain 
\begin{equation}
a\frac{d \beta}{d a} = -0.279(24)(^{+40}_{-64})
,\hspace{5mm}
a\frac{d \kappa_{ud}}{d a} = 0.00123(41)(^{+56}_{-68})
,\hspace{5mm}
a\frac{d \kappa_s}{d a} = 0.00046(26)(^{+42}_{-44})
\end{equation}
at our simulation point,
where the first brackets are for statistical errors, and the second brackets are for systematic errors estimated by the variation of the scale setting.

\section{Equation of state}
\label{sect:eos}

\begin{table}[tb]
\begin{tabular}{cccllll}
\hline
$N_t$ & $T$[MeV] & $N_{conf}$ & \hspace{3mm}$\langle S_{ud}^{\rm hopp}\rangle$ &
\hspace{3mm}$\langle S_{ud}^{\rm diag}\rangle$ & \hspace{3mm}$\langle S_s^{\rm hopp}\rangle$ &
\hspace{3mm}$\langle S_s^{\rm diag}\rangle$ \\
\hline
58 & --    &  390  &   -4.90487(46)  &  1.904649(79)    & -4.74878(44)
 & 1.909956(75)\\
\hline
16& 174 &  447 &   -0.00380(82)   &    -0.00065(13)   & -0.00271(80)
&    -0.00052(12)\\
14& 199 &  447 &   -0.0125(10)   &    -0.00182(17)   &    -0.01007(93)
&    -0.00153(16)\\
12& 232 &  495 &   -0.02987(88)   &    -0.00443(14)   &-0.02590(86)
&    -0.00394(14)\\
10& 279 &  287 &   -0.0448(11)   &    -0.00679(18)  &    -0.0422(12)
&    -0.00646(18)\\
8& 348 &  319 &   -0.0576(11)   &    -0.00885(15)   &    -0.0592(11)
&    -0.00898(15)\\
6& 464 &  159 &   -0.0850(15)   &    -0.01394(21)   &    -0.0947(15)
&    -0.01500(21)\\
4& 697 &    95 &   -0.3216(24)   &    -0.04966(38)   &  -0.3501(23)
&    -0.05266(35)\\
\hline
\end{tabular}
\caption{Quark contributions to the trace anomaly:
$
S_f^{\rm hopp} = (N_s^3 N_t)^{-1} 
\sum_{x,\mu}\mbox{Tr}^{(c,s)} \{(1-\gamma_\mu) U_{x,\mu} (D^{f})^{-1}_{x+\hat{\mu},x}
+(1+\gamma_\mu) U^\dagger_{x-\hat{\mu},\mu} (D^{f})^{-1}_{x-\hat{\mu},x}
\}
$
and
$
S_f^{\rm diag} =  (N_s^3 N_t)^{-1} 
\sum_{x,\mu>\nu}\mbox{Tr}^{(c,s)}\sigma_{\mu\nu} F_{\mu\nu} (D^{f})^{-1}_{x,x}
$.
In this table, the zero-temperature results ($N_t=58$) are raw expectation values, while the finite-temperature results are subtracted by the corresponding zero-temperature values.
Quark observables are measured every five trajectories after thermalization of 1000 trajectories, and their errors are estimated by adopting the bin size of 25 trajectories.
$N_{conf}$ is the number of configurations.}
\label{tab3}
\end{table}

With our lattice action (\ref{eq:gaction}) and (\ref{eq:qaction}), the trace anomaly $(\epsilon-3p)/T^4$  is given by
\begin{eqnarray}
\frac{\epsilon-3p}{T^4}&=& 
\frac{N_t^3}{N_s^3}
\left(
{a\frac{d\beta}{da}}
\left\langle
\frac{\partial S}{\partial\beta}
\right\rangle_{\rm \! sub}
+
{a\frac{d\kappa_{ud}}{da}}
\left\langle
\frac{\partial S}{\partial \kappa_{ud}}
\right\rangle_{\rm \! sub}
+{a\frac{d\kappa_s}{da}}
\left\langle
\frac{\partial S}{\partial \kappa_s}
\right\rangle_{\rm \! sub}
\right) 
\label{eq:tranom}
\end{eqnarray}
with
\begin{eqnarray}
\left\langle 
\frac{\partial S}{\partial \beta} 
\right\rangle_{\rm \! sub} &=&
-\left\langle
\sum_{x,\mu>\nu}c_0W^{1\times 1}_{\mu\nu}(x)
+\sum_{x,\mu,\nu}c_1W^{1\times 2}_{\mu\nu}(x)
\right\rangle_{\rm \! sub} 
\nonumber\\&&
+\; \frac{\partial c_{SW}}{\partial \beta}
\sum_{f=u,d,s}{\kappa_f
\left\langle 
\sum_{x,\mu>\nu}\mbox{Tr}^{(c,s)}\sigma_{\mu\nu}F_{\mu\nu}
(D^{f})^{-1}_{x,x}
\right\rangle_{\rm \! sub} },
\label{eq:dsdb}
\\
\left\langle 
\frac{\partial S}{\partial \kappa_f} 
\right\rangle_{\rm \! sub} &=&
N_f
\left(
\left\langle
\sum_{x,\mu}\mbox{Tr}^{(c,s)}
\{(1-\gamma_\mu)U_{x,\mu}(D^{f})^{-1}_{x+\hat{\mu},x}
+(1+\gamma_\mu)U^\dagger_{x-\hat{\mu},\mu}
(D^{f})^{-1}_{x-\hat{\mu},x}
\}
\right\rangle_{\rm \! sub} 
\right.
\nonumber\\&&
\left.
+\; c_{SW}
\left\langle
\sum_{x,\mu>\nu}\mbox{Tr}^{(c,s)}\sigma_{\mu\nu}F_{\mu\nu}
(D^{f})^{-1}_{x,x}
\right\rangle_{\rm \! sub}
\right),
\label{eq:dsdk}
\end{eqnarray}
where $N_f=2$ for $f=ud$ and 1 for $f=s$.
We evaluate the traces in (\ref{eq:dsdb}) and (\ref{eq:dsdk}) by the random noise method with complex U(1) random numbers \cite{Ejiri:2009hq}.
The number of noise vectors is 1 for each of the color and spinor indices.
Results of the quark contributions in (\ref{eq:dsdb}) and (\ref{eq:dsdk}) are summarized in Table~\ref{tab3}.

\begin{figure}[bt]
  \begin{center}
    \includegraphics[width=110mm]{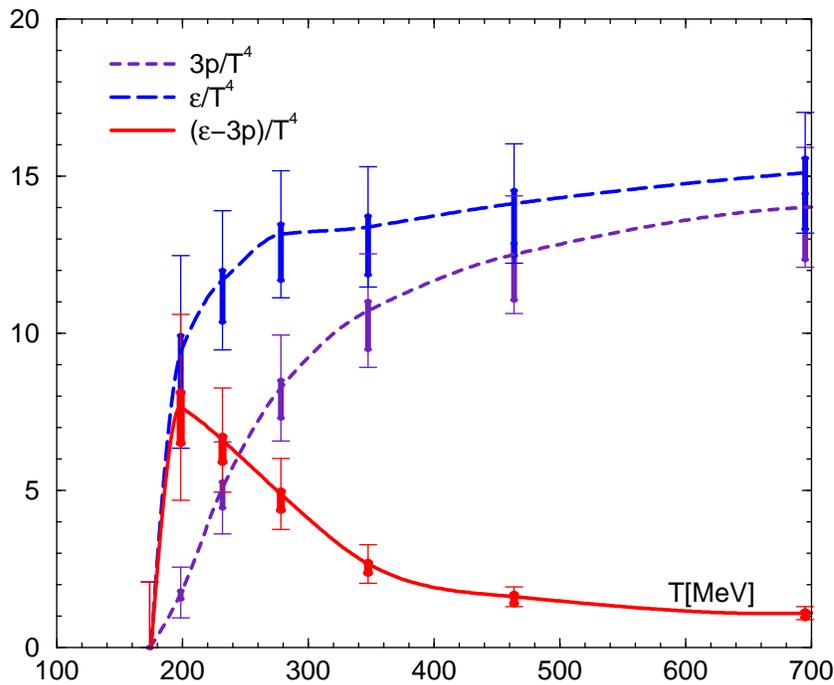}
    \caption{
    Trace anomaly $(\epsilon-3p)/T^4$, energy density $\epsilon/T^4$
    and pressure $3p/T^4$ in 2+1 flavor QCD. 
    The thin and thick vertical bars represent statistic and systematic errors, respectively.
    The curves are drawn by the Akima spline interpolation.
    }
    \label{fig:eos}
  \end{center}
\end{figure}

In Fig.~\ref{fig:eos}, the results of the trace anomaly (\ref{eq:tranom}) are shown by the solid curve.
The curve is drawn by the Akima spline interpolation \cite{akima1970}. 
The central values are the results using the beta functions with the scale
setting variable $am_\rho$, and vertical thin bars represent statistic
errors, in which the statistical errors of gauge and quark observables as well as those of the  beta functions are combined by the error propagation rule.
We repeat the calculation using the values of the beta functions adopting alternative scale setting variables to estimate the systematic error due to the beta function.
We find that the effect of the change of the scale setting variable partially cancels with each other among different beta functions in the trace anomaly. 
Resulting systematic errors are shown by thick vertical bars in Fig.~\ref{fig:eos}.
The systematic errors thus estimated are smaller than the statistical errors in this study.

We find that $(\epsilon-3p)/T^4$ is small at $T=174$ MeV but shows a peak at $T=199$ MeV and decreases towards higher $T$.
We note that the peak height of about 7 at $T=199$ MeV ($N_t = 14$) is roughly consistent with recent results of highly improved staggered quarks (obtained at $N_t = 6$--12) in the fixed-$N_t$ approach \cite{Borsanyi:2010cj,Bazavov:2010sb}.
The shape of $(\epsilon-3p)/T^4$ suggests that $T_{\rm pc}$ is located between 174 and 199 MeV.

Carrying out the $T$-integration (\ref{eq:Tintegral}) using the Akima spline interpolation for the trace anomaly, we obtain the pressure $p/T^4$ shown in Fig.~\ref{fig:eos}.
Here, we have chosen the starting point of the integration to be 
at $N_t=16$, where the trace anomaly vanishes within the statistical error.
The energy density $\epsilon/T^4$ is calculated by $p/T^4$ and $(\epsilon-3p)/T^4$.
To our knowledge, this is the first result for EOS in 2+1 flavor QCD with dynamical Wilson-type quarks.

In our previous test in quenched QCD, we confirmed that the choice of the interpolation procedure has only minor effects on the EOS \cite{Umeda:2008bd}.
Because the resolution in $T$ is coarser in the present study, we need to reexamine the influence of the interpolation procedures on the final values of the EOS.
The results are summarized in Appendix~\ref{app1}.
We find that the systematic errors due to the choice of the interpolation procedure are small in the EOS in comparison with the present statistical errors.

The overall large errors in $p/T^4$ and $\epsilon/T^4$ are mainly due to the large statistic error in $(\epsilon-3p)/T^4$ at $T\sim200$ MeV --- they propagate to higher $T$'s through the numerical integration.
The large statistic error in  $(\epsilon-3p)/T^4$ at $T\simle 200$ MeV is caused by the enhancement factor $N_t^4$ in (\ref{eq:tranom}) (note that $S$ is proportional to $N_t N_s^3$).
Although the central value is largely canceled by the zero-temperature subtraction procedure, the errors are magnified. 
We find that the statistical fluctuation is much larger in the gauge part than in the quark parts.
Note that the same difficulty exists also in the fixed-$N_t$ approach when we increase $N_t$ towards the continuum limit.
In the fixed-scale approach, because high statistics is required at very low temperatures only, the overall numerical cost will still be lower than that in the fixed-$N_t$ approach
when we try to keep a similar magnitude of discretization errors around the transition temperature.
In the present test, however, we stop at the current statistics and leave the task for the future  investigation at the physical point.

An additional source of errors in Fig.~\ref{fig:eos} is the spacing of the data points in $T$:
Because our lattice spacing $a$ is coarser than that of our previous study in quenched QCD \cite{Umeda:2008bd}, and also because $N_t$ is restricted to be even due to the CPS simulation code with the even-odd preconditioning, we cannot have the resolution as achieved in our previous study.
To improve the resolution in $T$, we need to develop a simulation code for odd $N_t$'s.
An alternative way out may be to combine results at different lattice spacing $a$.
Note that we can choose small values of $a$ in the fixed-scale approach. 
When $a$'s are well in the scaling region, results for physical observables as functions of $T$ should lie on the same curves for these $a$'s, but at different discrete points.
After confirming the insensitivity to a variation of $a$, we may combine the results at different $a$'s to more smoothly interpolate the data in $T$.
We leave the application of these methods to future studies of EOS at the physical point.

Besides the large errors, our EOS looks roughly consistent with recent results with highly improved staggered quarks near the physical point: 
The peak of the trace anomaly from the stout quarks locates at $T\approx 190$-200 MeV 
with a peak height of about 4.0 \cite{Borsanyi:2010cj}.
A preliminary result from the HISQ quarks gives a peak height of about 5.6 at $T\approx200$-220 MeV \cite{Bazavov:2010sb}.
We recall that our light quark masses are heavier than their physical values. 
The experience with improved staggered quarks suggests that the peak becomes higher as the light quark masses are increased (see, e.g., \cite{Borsanyi:2010cj}).

\section{Summary}
\label{sec:summary}

We calculated the EOS in 2+1 flavor QCD with improved Wilson quarks by adopting the fixed-scale approach \cite{Umeda:2008bd}, 
with which we vary $T$ without varying the system volume on a fine lattice.
As the first step towards the EOS with Wilson-type quarks in 2+1 flavor QCD, we made simulations at $m_\pi/m_\rho \simeq 0.63$, 
taking advantage of the fixed-scale approach to make use of high-precision configurations by the CP-PACS+JLQCD Collaboration at $T=0$ \cite{Ishikawa:2007nn}.
Although the light quark masses are still heavier than their physical values, 
our EOS looks roughly consistent with recent results with highly improved staggered 
quarks near the physical point \cite{Borsanyi:2010cj,Bazavov:2010sb}.

To extend the study towards the physical point, however, we found a couple of issues that need to be solved:
To obtain statistically accurate EOS at low temperatures, we need a large statistics proportional to $N_t^7$ (a power of $N_t$ is reduced due to the average over the lattice sites). 
This is, however, an unavoidable step to calculate observables suppressing discretization errors.
Another source of systematic errors in EOS is the limited resolution in $T$ due to the discrete variation of $N_t$ in the fixed-scale approach.
In the present study, because the lattice spacing $a$ is coarser than our previous quenched study, and because $N_t$ is limited to be even due to the simulation program set we have adopted, this seems to be non-negligible.
To improve the resolution in $T$, we need simulations at odd values of $N_t$ and a finer lattice spacing $a$.
An alternative way will be to combine results at different $a$'s, since we can choose five $a$'s with the fixed-scale approach; thus, after confirming that the discretization effects are sufficiently small in the observables under study, we may combine the results at different $a$'s to more smoothly interpolate in $T$.
We leave these trials to a forthcoming study with much lighter quarks, adopting the on-the-physical-point configurations by the PACS-CS Collaboration \cite{Aoki:2009ix}.

\section*{Acknowledgments}
We thank the members of the CP-PACS and JLQCD Collaborations for 
providing us with their high-statistics 2+1 flavor QCD configurations with improved Wilson quarks at $T=0$,
and the authors and maintainer of CPS++ \cite{CPS}, whose modified version is used in this paper.
This work is, in part, supported 
by Grants-in-Aid of the Japanese Ministry
of Education, Culture, Sports, Science and Technology, 
(No.\ 20340047,  
No.\ 22740168, 
No.\ 21340049 , 
No.\ 23540295).  
SA, SE and TH are supported by the Grant-in-Aid for Scientific Research on Innovative Areas
(No.\ 2004:20105001, No.\ 20105003, No.\ 2310576). 
This work is, in part, also supported by the Large Scale Simulation Program of the High Energy Accelerator Research Organization (KEK) No.\ 09/10-25 and No.\ 10-09.
HO is supported by the Japan Society for the Promotion of Science for
Young Scientists.

\appendix
\section{Comparison of interpolation procedures}
\label{app1}

\begin{figure}[bt] 
  \begin{center}
    \includegraphics[width=85mm]{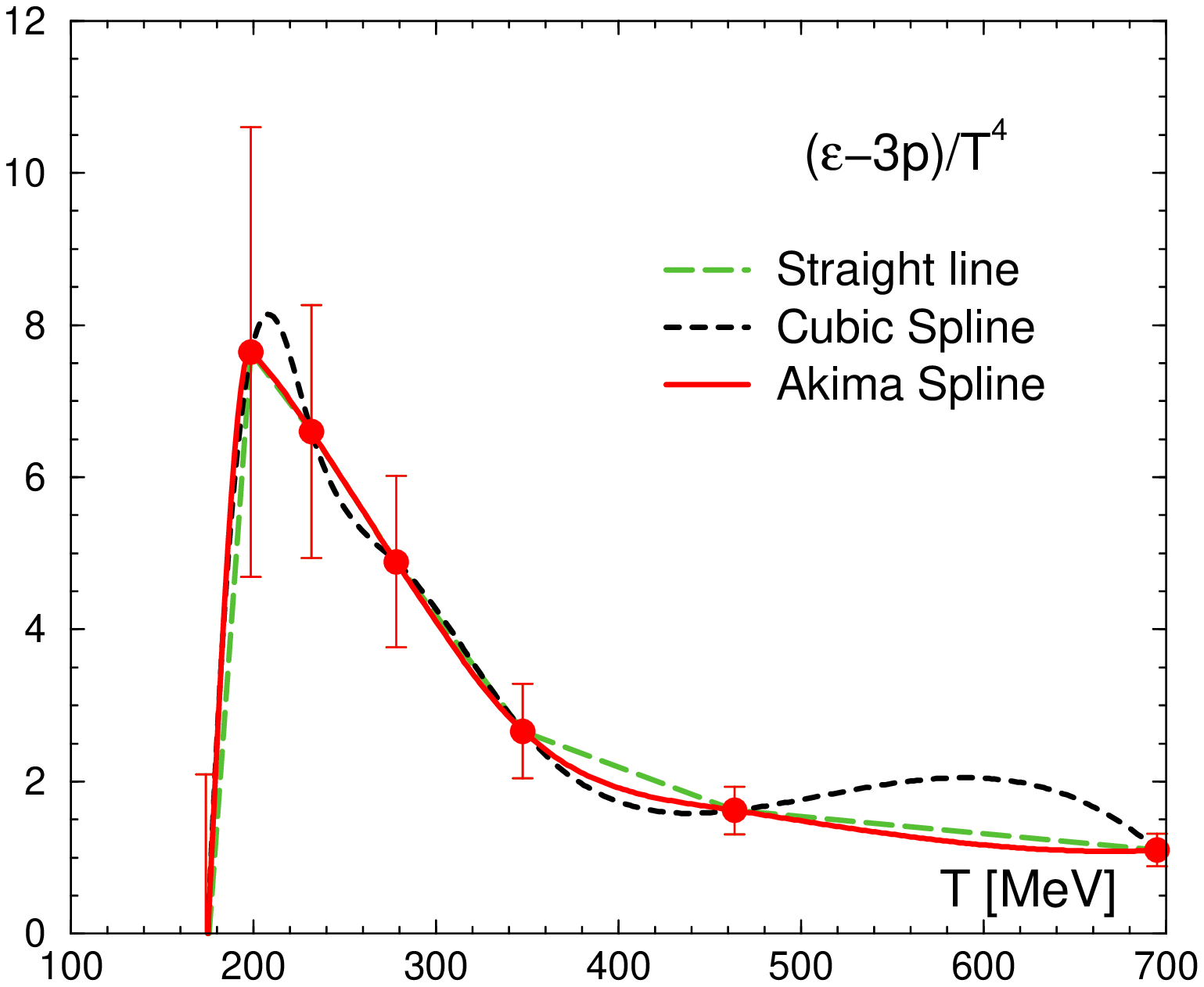}
    \hspace{5mm}
    \includegraphics[width=85mm]{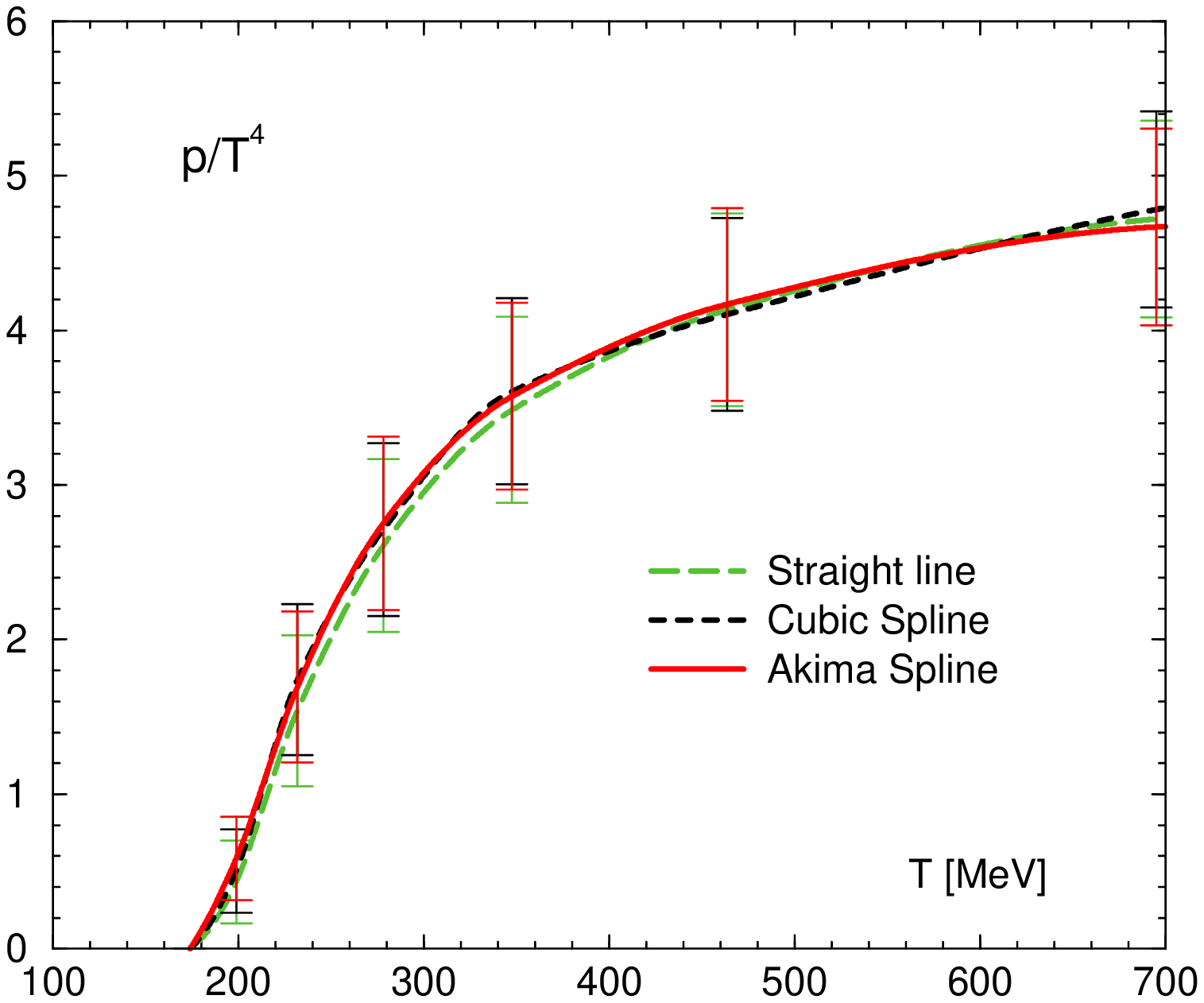}
    \caption{
    Straight line, cubic spline, and Akima spline interpolations for the trace anomaly and the pressure.}
    \label{fig:splines}
  \end{center}
\end{figure}

To carry out the $T$ integration given by (\ref{eq:Tintegral}), we need to interpolate the data of the trace anomaly at discrete values of $T$ corresponding to the discrete values of $N_t$.
In this appendix, we examine the interpolation procedures and their influences on the EOS with our data.

In the left panel of Fig.~\ref{fig:splines}, we apply three different interpolation procedures to our data of the trace anomaly.
Beta functions with the scale setting variable $am_\rho$ are adopted. 
The long-dotted line, dotted line, and solid line represent the results of straight line, cubic spline, and Akima spline \cite{akima1970} interpolations, respectively. 
In our previous study in quenched QCD, we have adopted the cubic spline interpolation \cite{Umeda:2008bd}.
With our present data, however, we find the oscillatory interpolation curve by the cubic spline interpolation.
This is due to the coarseness of the present data points --- data are available only at even values of $N_t$.
The cubic spline is not stable for data sets with sharp variations. 

In such cases, the Akima spline interpolation \cite{akima1970} is widely adopted.
The Akima spline is a combination of local cubic polynomials and is known to suppress such oscillatory behavior around sharp variations. 
From Fig.~\ref{fig:splines}, we find that the Akima spline leads to a more natural curve smoothly following the data points.
Therefore, we adopt the Akima spline interpolation in this study.


To estimate the systematic error due to the choice of the interpolation procedure in the EOS, 
we perform the $T$ integration with these interpolations.
The results for the pressure are shown in the right panel of Fig.~\ref{fig:splines}.
The strong oscillation of the interpolation curve from the cubic spline is averaged over through the integration, and the results of $p/T^4$ are well consistent for all three interpolations.
We thus conclude that the systematic error in the EOS due to the choice of the interpolation procedure is much smaller than the statistical errors.



\begin{thebibliography}{99}

\bibitem{Hirano:2008hy}
  T.~Hirano, N.~van der Kolk, and A.~Bilandzic,
  Lect.\ Notes Phys.\  {\bf 785}, 139 (2010)
  [arXiv:0808.2684 [nucl-th]].

\bibitem{Levkova:2006gn} 
  L.~Levkova, T.~Manke and R.~Mawhinney,
  Phys.\ Rev.\ D {\bf 73}, 074504 (2006)
  [hep-lat/0603031].

\bibitem{Umeda:2008bd}
  T.~Umeda, S.~Ejiri, S.~Aoki, T.~Hatsuda, K.~Kanaya, Y.~Maezawa, and H.~Ohno 
  [WHOT-QCD Collaboration],
  Phys.\ Rev.\  D {\bf 79}, 051501 (2009)
  [arXiv:0809.2842 [hep-lat]].

\bibitem{Maynard:2010wi}
  C.M.~Maynard,
  PoS LAT2009, 020 (2009)
  [arXiv:1001.5207 [hep-lat]].

\bibitem{Borsanyi:2010cj}
  S.~Borsanyi, G.~Endrodi, Z.~Fodor, A.~Jakovac, S.D.~Katz, S.~Krieg, C.~Ratti, and K.K.~Szabo,
  JHEP 1011, 077 (2010)
  [arXiv:1007.2580 [hep-lat]].

\bibitem{Bazavov:2010sb}
  A.~Bazavov and P.~Petreczky  [HotQCD Collaboration],
  J.\ Phys.\ Conf.\ Ser.\  {\bf 230}, 012014 (2010)
  [arXiv:1005.1131 [hep-lat]].

\bibitem{DIK2010}
 V.G.~Bornyakov, R.~Horsley, S.M.~Morozov, Y.~Nakamura, M.I.~Polikarpov, P.E.L.~Rakow, G.~Schierholz, and T.~Suzuki,
  Phys.\ Rev.\ D {\bf 82}, 014504 (2010)
  [arXiv:0910.2392 [hep-lat]].
  
\bibitem{Conf10}
 B.B.~Brandt, O.~Philipsen, H.~Wittig, and L.~Zeidlewicz,
  AIP Conf.\ Proc.\  {\bf 1343}, 516 (2011)
  [arXiv:1011.6172 [hep-lat]].
  
\bibitem{tmfT11}
 F.~Burger, E.-M.~Ilgenfritz, M.~Kirchner, M.~P.~Lombardo, M.~Muller-Preussker, O.~Philipsen, C.~Urbach, and L.~Zeidlewicz,
  arXiv:1102.4530 [hep-lat].

\bibitem{BWLat11}
 S.~Borsanyi, Z.~Fodor, C.~Hoelbling, S.D.~Katz, S.~Krieg, D.~Nogradi, B.~C.~Toth, and K.K.~Szabo,
  arXiv:1111.3500 [hep-lat].

\bibitem{AliKhan:2000}
 A.~Ali Khan, S.~Aoki, R.~Burkhalter, S.~Ejiri, M.~Fukugita, S.~Hashimoto, N.~Ishizuka, Y.~Iwasaki, K.~Kanaya, T.~Kaneko, Y.~Kuramashi, T.~Manke, K.I.~Nagai, M.~Okamoto, M.~Okawa, A.~Ukawa, and T.~Yoshi\'e 
  [CP-PACS Collaboration],
  Phys.\ Rev.\ D {\bf 63}, 034502 (2001)
  [hep-lat/0008011].

\bibitem{AliKhan:2001ek}
  A.~Ali Khan , S.~Aoki, R.~Burkhalter, S.~Ejiri, M.~Fukugita, S.~Hashimoto, N.~Ishizuka, Y.~Iwasaki, K.~Kanaya, T.~Kaneko, Y.~Kuramashi, T.~Manke, K.-I.~Nagai, M.~Okamoto, M.~Okawa, H.P.~Shanahan, Y.~Taniguchi, A.~Ukawa, and T.~Yoshi\'e 
  [CP-PACS Collaboration],
  Phys.\ Rev.\  D {\bf 64}, 074510 (2001)
  [arXiv:hep-lat/0103028].

\bibitem{Aoki:2005et}
  S.~Aoki M.~Fukugita, S.~Hashimoto, K.-I.~Ishikawa, N.~Ishizuka, Y.~Iwasaki, K.~Kanaya, T.~Kaneko, Y.~Kuramashi, M.~Okawa, S.~Takeda, Y.~Taniguchi, N.~Tsutsui, A.~Ukawa, N.~Yamada, and T.~Yoshi\'e 
  [CP-PACS and JLQCD Collaborations],
  Phys.\ Rev.\  D {\bf 73}, 034501 (2006).
  [arXiv:hep-lat/0508031].

\bibitem{Ishikawa:2007nn}
  T.~Ishikawa, S.~Aoki M.~Fukugita, S.~Hashimoto, K.-I.~Ishikawa, N.~Ishizuka, Y.~Iwasaki, K.~Kanaya, T.~Kaneko, Y.~Kuramashi, M.~Okawa, Y.~Taniguchi, N.~Tsutsui, A.~Ukawa, N.~Yamada, and T.~Yoshi\'e
  [CP-PACS and JLQCD Collaborations],
  Phys.\ Rev.\  D {\bf 78}, 011502 (2008)
  [arXiv:0704.1937 [hep-lat]].
  
\bibitem{Kanaya:2009nf}
  K.~Kanaya, T.~Umeda, S.~Aoki, S.~Ejiri, T.~Hatsuda, N.~Ishii, Y.~Maezawa, and H.~Ohno  
  [WHOT-QCD Collaboration],
  Nucl.\ Phys.\  A {\bf 830}, 801C (2009)
  [arXiv:0907.4205 [hep-lat]];
  K.~Kanaya, S.~Aoki, H.~Ohno, T.~Umeda, S.~Ejiri, T.~Hatsuda, N.~Ishii, and Y.~Maezawa 
  [WHOT-QCD Collaboration],
  PoS {\bf LAT2009}, 190 (2009)
  [arXiv:0910.5284 [hep-lat]].
  
\bibitem{Umeda:2010ye}
 T.~Umeda, S.~Aoki, K.~Kanaya, H.~Ohno, S.~Ejiri, T.~Hatsuda, and Y.~Maezawa 
  [WHOT-QCD Collaboration],
  PoS LATTICE {\bf 2010}, 218 (2010)
  [arXiv:1011.2548 [hep-lat]].

\bibitem{Engels:1990vr}
  J.~Engels, J.~Fingberg, F.~Karsch, D.~Miller and M.~Weber,
  Phys.\ Lett.\  B {\bf 252}, 625 (1990).

\bibitem{Sheikholeslami:1985ij}
  B.~Sheikholeslami and R.~Wohlert,
  Nucl.\ Phys.\  B {\bf 259}, 572 (1985).

\bibitem{Iwasaki}
 Y.~Iwasaki,
 Nucl.\ Phys.\ B258, 141 (1985); University of Tsukuba Report No. UTHEP-118, (1983) 
  [arXiv:1111.7054 [hep-lat]].

\bibitem{Aoki:2009sf}
 S.~Aoki, K.-I.~Ishikawa, N.~Ishizuka, T.~Izubuchi, D.~Kadoh, K.~Kanaya, Y.~Kuramashi, K.~Murano, Y.~Namekawa, M.~Okawa, Y.~Taniguchi, A.~Ukawa, N.~Ukita, and T.~Yoshi\'e  
 [PACS-CS Collaboration],
  JHEP {\bf 0910}, 053 (2009)
  [arXiv:0906.3906 [hep-lat]].

\bibitem{Aoki:2009pp}
 S.~Aoki, K.-I.~Ishikawa, N.~Ishizuka, T.~Izubuchi, D.~Kadoh, K.~Kanaya, Y.~Kuramashi, Y.~Namekawa, M.~Okawa, Y.~Taniguchi, A.~Ukawa, N.~Ukita, and T.~Yoshi\'e 
  [PACS-CS Collaboration],
  Phys.\ Rev.\ D {\bf 79}, 034503 (2009)
  [arXiv:0807.1661 [hep-lat]].
 
\bibitem{Ishikawa:2009su}
 K.-I.~Ishikawa, N.~Ishizuka, T.~Izubuchi, D.~Kadoh, K.~Kanaya, Y.~Kuramashi, Y.~Namekawa, M.~Okawa, Y.~Taniguchi, A.~Ukawa, N.~Ukita, and T.~Yoshi\'e  
 [PACS-CS Collaboration],
  Phys.\ Rev.\ D {\bf 80}, 054502 (2009)
  [arXiv:0905.0962 [hep-lat]].

\bibitem{Aoki:2009ix}
  S.~Aoki, K.-I.~Ishikawa, N.~Ishizuka, T.~Izubuchi, D.~Kadoh, K.~Kanaya, Y.~Kuramashi, Y.~Namekawa, M.~Okawa, Y.~Taniguchi, A.~Ukawa, N.~Ukita, T.~Yamazaki, and T.~Yoshi\'e 
  [PACS-CS Collaboration],
  Phys.\ Rev.\  D {\bf 81}, 074503 (2010)
  [arXiv:0911.2561 [hep-lat]].
  
\bibitem{Maezawa:2009di}
  Y.~Maezawa, S.~Aoki, S.~Ejiri, T.~Hatsuda, K.~Kanaya, H.~Ohno, and T.~Umeda
  [WHOT-QCD Collaboration],
  PoS LAT2009, 165 (2009)
  [arXiv:0911.0254 [hep-lat]].

\bibitem{CPS}
Columbia Physics System (CPS), http://qcdoc.phys.columbia.edu/cps.html

\bibitem{Ejiri:2009hq}
  S.~Ejiri, Y.~Maezawa, N.~Ukita, S.~Aoki, T.~Hatsuda, N.~Ishii, K.~Kanaya, and T.~Umeda
   [WHOT-QCD Collaboration],
  Phys.\ Rev.\  D {\bf 82}, 014508 (2010)
  [arXiv:0909.2121 [hep-lat]].

\bibitem{Maezawa2011}
Y.~Maezawa, T.~Umeda, S.~Aoki, S.~Ejiri, T.~Hatsuda, K.~Kanaya, and H.~Ohno,
  arXiv:1112.2756 [hep-lat].

\bibitem{RBCB2008}
  M.~Cheng, N.H.~Christ, S.~Datta, J.~van der Heide, C.~Jung, F.~Karsch, O.~Kaczmarek and E.~Laermann,
  Phys.\ Rev.\ D {\bf 77}, 014511 (2008)
  [arXiv:0710.0354 [hep-lat]].

\bibitem{Luescher}
M.~L\"uscher, K.~Symanzik, and P.~Weisz,
 Nucl.\ Phys.\ B {\bf 173}, 365 (1980);
M.~L\"uscher,
 Nucl.\ Phys.\ B {\bf 180}, 317 (1981); 
M.~L\"uscher, and P.~Weisz,
  JHEP {\bf 0207}, 049 (2002)
  [arXiv:hep-lat/0207003].

\bibitem{RBCB2010}
 M.~Cheng, S.~Ejiri, P.~Hegde, F.~Karsch, O.~Kaczmarek, E.~Laermann, R.D.~Mawhinney, and C.~Miao,
  Phys.\ Rev.\ D {\bf 81}, 054504 (2010)
  [arXiv:0911.2215 [hep-lat]].



\bibitem{Umeda:2008kz}
  T.~Umeda, S.~Ejiri, S.~Aoki, T.~Hatsuda, K.~Kanaya, Y.~Maezawa, and H.~Ohno 
  [WHOT-QCD Collaboration],
  PoS LAT2008, 174 (2008)
  [arXiv:0810.1570 [hep-lat]].

\bibitem{RBCBi2007}
M.~Cheng, N.H.~Christ, S.~Datta, J.~van der Heide, C.~Jung, F.~Karsch, O.~Kaczmarek, and E.~Laermann,
  Phys.\ Rev.\ D {\bf 74}, 054507 (2006)
  [hep-lat/0608013].

\bibitem{akima1970}
H.~Akima, J.\ Assoc.\ Comp.\ Mach.\ 17, 589 (1970).


\end{thebibliography}
\end{document}